\documentclass[twoside,12pt]{report}
\usepackage{graphicx}
\usepackage{amssymb, amsmath, amsxtra}
\usepackage{epsfig,subfigure}

\usepackage{latexsym}
\usepackage{bm}

 \usepackage{ntheorem}
           \theoremstyle{plain}
                      {\theorembodyfont{\rmfamily}
                      \theoremseparator{.}
                       
           \newtheorem{thm}{Theorem}[section]
           \theoremstyle{plain}
           
           \theoremstyle{plain} 
           \theoremstyle{plain}

           \theoremstyle{plain}
           \newtheorem{rem}{Remark}[section]
            }

\begin{document}
\begin{center}
{\Large \textbf{GROUP ANALYSIS OF NONLINEAR}\\[1ex] \textbf{INTERNAL WAVES IN
OCEANS}}\\[1.5ex]
 {\Large \textbf{II: The symmetries and rotationally invariant solution}}\footnote{Published
 in \textit{Archives of ALGA}, vol. 6, 2009, pp. 45-54.}\\[1ex]
 {\sc Nail H. Ibragimov}\\
 Department of Mathematics and Science, Blekinge Institute
 of Technology,\\ 371 79 Karlskrona, Sweden\\[.5ex]
 {\sc Ranis N. Ibragimov}\\
Department of Mathematics, Research and Support Center for Applied
Mathematical Modeling (RSCAMM),\\
 New Mexico Institute of Mining and Technology,\\
 Socorro, NM, 87801 USA\\[.5ex]
  {\sc Vladimir F. Kovalev}\\
Institute of Mathematical Modelling, Russian Academy of Sciences,\\
Miusskaya Sq. 4a, Moscow 124047, Russia
 \end{center}

 \noindent
\textbf{Abstract.} The maximal group of Lie point symmetries
 of a system of nonlinear equations used in geophysical fluid dynamics is presented.
 The Lie algebra of this group is infinite-dimensional and involves three
 arbitrary functions of time. The invariant solution under the rotation and dilation is
 constructed. Qualitative analysis of the invariant solution is
 provided and the energy of this solution is presented.\\[2ex]


  \section{Introduction}
 \setcounter{equation}{0}
This is a continuation of the paper \cite{ibr-ibr09b}. We present
here the Lie algebra of the maximal group of Lie point symmetries
for system nonlinear equations
 \begin{align}
 \Delta \psi_t - g \rho_x - f v_z & =
 \psi_x \Delta \psi_z - \psi_z \Delta \psi_x\,, \label{iik.eq1a}\\[1ex]
 v_t + f \psi_z & = \psi_x v_z - \psi_z v_x\,, \label{iik.eq2a}\\[1ex]
 \rho_t + \frac{N^2}{g}\, \psi_x  & = \psi_x \rho_z - \psi_z \rho_x \label{iik.eq3a}
 \end{align}
 used in geophysical fluid dynamics, e.g. for
 investigating internal waves in
 uniformly stratified incompressible fluids (oceans).
 Here $g, f, N$ are constants and $\Delta$ is the
 two-dimensional Laplacian:
 $$
 \Delta = D_x^2 + D_z^2\,\cdot
 $$

 \section{Symmetries}
 \setcounter{equation}{0}

 \subsection{General case}

The point symmetries of Eqs. (\ref{iik.eq1a})-(\ref{iik.eq3a}) have
been computed with the help of DIMSYM 2.3 package. The maximal
admitted Lie point transformation group is infinite for arbitrary
constants $f$ and $N.$ If $f \not= 0,$ the group is generated by the
infinite-dimensional Lie algebra spanned by the following operators:
\begin{align}
  &
 X_1= \frac{\partial}{\partial v}  \,, \quad
 X_2= \frac{\partial}{\partial \rho}\,, \quad
 X_3= a(t) \frac{\partial}{\partial \psi}\,, \quad X_4=
 \frac{\partial}{\partial t} \,,\notag \\[.5ex]
  &
  X_5= b(t) \left[\frac{\partial}{\partial x} - f\,\frac{\partial}{\partial v}\right] +
       b'(t) z\,\frac{\partial }{\partial \psi}\,, \notag\\[.5ex]
   &
 X_6= c(t) \left[\frac{\partial }{\partial  z} + \frac{N^2}{g}\,
  \frac{\partial}{\partial \rho}\right] -
  c'(t) x\,\frac{\partial}{\partial \psi}\,, \label{kov} \\[.5ex]
  &
 X_7= x \frac{\partial}{\partial x}  +
      z \frac{\partial}{\partial z} +
      v \frac{\partial}{\partial v} +
      \rho \frac{\partial}{\partial \rho}  +
      2 \psi \frac{\partial}{\partial \psi}\,, \notag\\[.5ex]
  &
 X_8= t \frac{\partial }{\partial  t}  +
      2 x \frac{\partial }{\partial  x}  +
      2 z \frac{\partial }{\partial  z}  +
      3 \psi \frac{\partial }{\partial  \psi} -
      2 f x \frac{\partial}{\partial v} +
    2 \frac{N^2}{g}\, z \frac{\partial}{\partial \rho} \,, \notag\\[.5ex]
    &
 X_9= z \frac{\partial}{\partial x} -
      x \frac{\partial}{\partial z} -
      \frac{1}{f}\,\big[g \rho + (f^2 - N^2) z\big]
      \frac{\partial}{\partial v}
     + \frac{1}{g}\,\big[fv + (f^2 - N^2) x \big]
       \frac{\partial}{\partial \rho}\,\cdot\notag
\end{align}
Here $a(t)$, $b(t)$ and $c(t)$ are arbitrary functions of time $t$.
 \begin{rem}
 The presence of the arbitrary functions $a(t), b(t), c(t)$ in the
 symmetry Lie algebra is a characteristic property of
 incompressible fluids (\cite{buch71}, see also \cite{akpr94}). Namely, the operator $X_3$ generates the
 group transformation $\bar \psi = \psi + \varepsilon_3\,a(t)$ of the stream
 function $\psi,$ where $\varepsilon_3$ is the group parameter.
 The invariance of fluid flows under this transformation is quite obvious
 because the velocity vector $(\psi_z, v, - \psi_x)$ is invariant under this transformation.
 The operators $X_5, X_6$ express the invariance
 under the generalization $\bar x = x + \varepsilon_5\,b(t), \ \bar z = z + \varepsilon_6\,c(t)$
 of the coordinate translations and the Galilean transformations.
 They provide a \textit{generalized relativity principle} for the Euler
 equations in terms of  conservation laws (see \cite{ibr83}, Section
 25.3).
 \end{rem}

 \subsection{The case $\bm{f = 0}$}

% \begin{rem}
 In order to include the special case $f =0,$ we multiply the
 operator $X_9$ by the constant $f$ and consider the operator
 $$
X'_9= f \Big[z \frac{\partial}{\partial x} -
      x \frac{\partial}{\partial z}\Big] -
      \big[g \rho + (f^2 - N^2) z\big]
      \frac{\partial}{\partial v}
     + \frac{f}{g}\,\big[fv + (f^2 - N^2) x \big]
       \frac{\partial}{\partial \rho}\,\cdot
 $$
 Then we let $f =0$ and obtain the operator
  $$
 X'_9 = - \big[g \rho  - N^2\, z\big]
      \frac{\partial}{\partial v}
 $$
 admitted by Eqs. (\ref{iik.eq1a})-(\ref{iik.eq3a}) with $f =0.$ The solution
 of the determining equations shows that $X'_9$ is a particular
 case of a more general symmetry involving an arbitrary function
 of two variables. Namely, the system (\ref{iik.eq1a})-(\ref{iik.eq3a}) with $f =0$
 admits the infinite-dimensional Lie algebra spanned by the following
operators:
\begin{equation}
 \label{kov2}
\begin{aligned}
  &
 X_1= h(v, g \rho  - N^2\, z)\,\frac{\partial}{\partial v}\,, \quad
 X_2= \frac{\partial}{\partial \rho}\,, \quad
 X_3= a(t) \frac{\partial}{\partial \psi}\,, \quad X_4=
 \frac{\partial}{\partial t} \,, \\[1ex]
  &
  X_5= b(t) \frac{\partial}{\partial x} +
       b'(t) z\,\frac{\partial }{\partial \psi}\,, \\ &
 X_6= c(t) \left[\frac{\partial }{\partial  z} + \frac{N^2}{g}\,
  \frac{\partial}{\partial \rho}\right] -
  c'(t) x\,\frac{\partial}{\partial \psi}\,, \\[1ex]
  &
 X_7= x \frac{\partial}{\partial x}  +
      z \frac{\partial}{\partial z} +
      v \frac{\partial}{\partial v} +
      \rho \frac{\partial}{\partial \rho}  +
      2 \psi \frac{\partial}{\partial \psi}\,, \\ &
 X_8= t \frac{\partial }{\partial  t}  +
      2 x \frac{\partial }{\partial  x}  +
      2 z \frac{\partial }{\partial  z}  +
      3 \psi \frac{\partial }{\partial  \psi} +
    2 \frac{N^2}{g}\, z \frac{\partial}{\partial \rho}\,,\\
\end{aligned}
\end{equation}
where $h(v, g \rho  - N^2\, z)$ is an arbitrary function of two
variables. The operator $X_1$ in (\ref{kov}) is obtained from the
operator $X_1$ in (\ref{kov2}) by taking $h = 1.$
% \end{rem}

% \begin{rem}
% The presence of the arbitrary functions $a(t), b(t), c(t)$ in the
% symmetry Lie algebra is a characteristic property of
% incompressible fluids (\cite{buch71}, see also \cite{akpr94}). Namely, the operator $X_3$ generates the
% group transformation $\bar \psi = \psi + \varepsilon_3\,a(t)$ of the stream
% function $\psi,$ where $\varepsilon_3$ is the group parameter.
% The invariance of fluid flows under this transformation is quite obvious
% because the velocity vector $(\psi_z, v, - \psi_x)$ is invariant under this transformation.
% The operators $X_5, X_6$ express the invariance
% under the generalization $\bar x = x + \varepsilon_5\,b(t), \ \bar z = z + \varepsilon_6\,c(t)$
% of the coordinate translations and the Galilean transformations.
% They provide a \textit{generalized relativity principle} for the Euler
% equations in terms of  conservation laws (see \cite{ibr83}, Section
% 25.3).
% \end{rem}

 \section{Invariant solution based on rotations and dilations}
 \setcounter{equation}{0}

 \subsection{The invariants}

 We will investigate here the invariant solutions with respect to
 the dilations and rotations with the generators $X_7$ and $X_9.$
 Let us introduce the notation
 \begin{equation}
 \label{Sph.eq4b}
 v_* = f v, \quad u = g \rho, \quad \alpha = f^2 - N^2
 \end{equation}
 and write the operators $X_7, \ X_9$ in the form
 \begin{equation}
 \label{Sph.eq4c}
 \begin{split}
   &
 X_7= x \frac{\partial}{\partial x}  +
      z \frac{\partial}{\partial z} +
      u \frac{\partial}{\partial u} +
      v_* \frac{\partial}{\partial v_*}  +
      2 \psi \frac{\partial}{\partial \psi}\,, \\
 & X_9= z \frac{\partial}{\partial x} -
      x \frac{\partial}{\partial z}
      + \big(v_* + \alpha\, x \big)
       \frac{\partial}{\partial u}
      - \big(u + \alpha\, z\big)
      \frac{\partial}{\partial v_*}\,\cdot
 \end{split}
 \end{equation}
 The operators (\ref{Sph.eq4c}) coincide with the operators (3.17) from
 \cite{ibr09},
 \begin{align}
 & X_1 = x \frac{\partial}{\partial x} + y \frac{\partial}{\partial y}
 + u \frac{\partial}{\partial u} + v \frac{\partial}{\partial v}
 + k w \frac{\partial}{\partial w}\,,\notag\\[1.5ex]
 & X_2 = y \frac{\partial}{\partial x} - x \frac{\partial}{\partial y}
 + (v + \alpha x + \beta y) \frac{\partial}{\partial u}
 - (u - \beta x + \alpha y) \frac{\partial}{\partial
 v}\,,\notag
 \end{align}
 with $k = 2$ and $\beta = 0$ upon identifying $v$ with $v_*$ and
 $y$ with $z.$ Hence, a basis of invariants for the operators (\ref{Sph.eq4c})
 contains the time $t$ and the invariants (3.20) from \cite{ibr09}
 which have now the form
 \begin{align}
  & J_1 = \frac{1}{x^2 + z^2}\, \left(x u + z v_* + \alpha xz\right)\,,\notag \\[1ex]
 & J_2 = \frac{1}{x^2 + y^2}\, \left(x v_* - z u  +
 \frac{\alpha}{2}(x^2 - z^2)\right)\,,\notag \\[1ex]
 & J_3 = \frac{\psi}{x^2 + y^2}\,\cdot\notag
 \end{align}
 It is more convenient for our purposes to use, instead of these
 invariants, the equivalent equations (3.19) from \cite{ibr09}
 which are written now as follows:
 \begin{equation}
 \label{Sph.eq4d}
 \begin{split}
 & u =  J_1\,x
 - \left(J_2 + \frac{\alpha}{2}\right)\,z\,, \\[1ex]
 & v_* = J_1\,z
 + \left(J_2 - \frac{\alpha}{2}\right)\,x\,, \\[1ex]
 & \psi = (x^2 + z^2)\, J_3.
 \end{split}
 \end{equation}

 \subsection{Candidates for the invariant solution}

 Knowledge of a symmetry algebra allows one to obtain particular exact solutions
 to differential equations in question. These kind of solutions were considered by
 S. Lie \cite{lie95}. They are known today as group invariant solutions
 (briefly \textit{invariant solutions}) and widely used in the modern literature,
 particularly in investigating nonlinear differential equations.
 %(see, e.g., \cite{ovs58, ovs62, ovs78}, \cite{ibr66}, \cite{blu-col74}, \cite{olv86},
 %\cite{blu-kum89}, \cite{ibr94, ibr95, ibr96}, \cite{akpr94}, \cite{can02}, \cite{blu-anc02}).

 The general form of regular invariant solutions is obtained from Eqs. (\ref{Sph.eq4d})
 by setting
 $$
 J_1 = R(t), \quad J_2 = V(t), \quad J_3 = \phi(t)
 $$
 with undetermined functions $R(t), \ V(t), \ \phi(t).$
 Invoking the notation (\ref{Sph.eq4b}) we arrive at the
 following   general form of candidates for the invariant solution
 with respect to
 the dilations and rotations with the generators $X_7$ and $X_9$
 from (\ref{kov}):
 \begin{equation}
 \label{Sph.eq4e}
 \begin{split}
 & v = \frac{1}{f} \left[R(t)\,z + V(t)\,x
 + \frac{N^2 - f^2}{2}\,x\right], \\[1ex]
 & \rho = \frac{1}{g} \left[R(t)\,x - V(t)\,z
 + \frac{N^2 - f^2}{2}\,z\right], \\[1ex]
 & \psi = (x^2 + z^2)\, \phi(t).
 \end{split}
 \end{equation}

 \begin{rem}
 Solving the Lie equations for the operator $X_9$ from (\ref{Sph.eq4c}) and using the notation (\ref{Sph.eq4b}),
 one can verify that the operator $X_9$ from (\ref{kov}),
 $$
  X_9= z \frac{\partial}{\partial x} -
      x \frac{\partial}{\partial z} -
      \frac{1}{f}\,\big[g \rho + (f^2 - N^2) z\big]
      \frac{\partial}{\partial v}
     + \frac{1}{g}\,\big[fv + (f^2 - N^2) x \big]
       \frac{\partial}{\partial \rho}\,,
  $$
  generates the following one-parameter transformation group with the parameter
  $\varepsilon:$
 \begin{equation}
 \label{Sph.eq4a9}
 \begin{split}
 & \bar x = x \cos \varepsilon + z \sin \varepsilon, \quad
 \bar z = z \cos \varepsilon - x \sin \varepsilon, \\[1ex]
 & g \bar \rho = g \rho \cos \varepsilon + f v \sin \varepsilon
  - (N^2 - f^2)\, x \sin \varepsilon, \\[1ex]
 & f \bar v = f v \cos \varepsilon - g \rho \sin \varepsilon
 + (N^2 - f^2)\, z \sin \varepsilon,\\[1ex]
 & \bar t = t, \quad \bar \psi = \psi.
 \end{split}
 \end{equation}
 One can verify by inspection that the transformations (\ref{Sph.eq4a9}) leave invariant Eqs.
 (\ref{Sph.eq4e}):
 \begin{align}
 & \bar v = \frac{1}{f} \left[R(t)\,\bar z + V(t)\,\bar x
 + \frac{N^2 - f^2}{2}\,\bar x\right],\notag \\[1ex]
 & \bar \rho = \frac{1}{g} \left[R(t)\,\bar x - V(t)\,\bar z
 + \frac{N^2 - f^2}{2}\,\bar z\right],\notag \\[1ex]
 & \bar \psi = (\bar x^2 + \bar z^2)\, \phi(t).\notag
 \end{align}
 \end{rem}

 \subsection{Construction of the invariant solution}

 It remains to determine the functions $R(t), \ V(t), \ \phi(t)$ by
 substituting the expressions (\ref{Sph.eq4e}) for
 $\rho,\ v,\ \psi$ in Eqs. (\ref{iik.eq1a})-(\ref{iik.eq3a}).

 Differentiating (\ref{Sph.eq4e}) we obtain:
 \begin{align}
 \label{Sph.eq4f}
 & v_t = \frac{1}{f}\left[R'\,z + V'\,x\right], \quad
 v_x = \frac{1}{f} \left[\frac{N^2 - f^2}{2} + V\right], \quad
 v_z = \frac{1}{f}\,R,\notag\\[1.5ex]
 & \rho_t =\frac{1}{g} \left[R'\,x - V'\,z\right], \quad
 \rho_x =\frac{1}{g}\, R,\quad
 \rho_z =\frac{1}{g} \left[\frac{N^2 - f^2}{2} - V\right],\\[1.5ex]
 & \psi_t = (x^2 + z^2)\, \phi', \quad \psi_x = 2 x\, \phi, \quad
 \psi_z = 2 z\, \phi, \quad \Delta \psi_t = 4 \phi'.\notag
 \end{align}
 Substitution of (\ref{Sph.eq4f}) in Eqs. (\ref{iik.eq1a})-(\ref{iik.eq3a}) yields:
 \begin{align}
 & 2 \phi' - R = 0,\label{Sph.eq1A}\\[1ex]
 & [V' - 2 R \phi]\, x + [R' + 2 V\phi + (N^2 + f^2)\phi]\, z= 0, \label{Sph.eq1B}\\[1ex]
 & [R' + 2 V\phi + (N^2 + f^2)\phi]\, x - [V' - 2 R \phi]\,z = 0.\label{Sph.eq1C}
 \end{align}
 Since $V, R, \phi$ depend only on $t,$ Eq. (\ref{Sph.eq1B})
 implies that
 \begin{equation}
 \label{Sph.eq1Ba}
 V' - 2 R \phi = 0
 \end{equation}
 and
 \begin{equation}
 \label{Sph.eq1Bb}
 R' + 2 V\phi + (N^2 + f^2)\phi = 0.
 \end{equation}
 Eq. (\ref{Sph.eq1C}) is satisfied due to Eqs. (\ref{Sph.eq1Ba}),
 (\ref{Sph.eq1Bb}). Hence, Eqs. (\ref{iik.eq1a})-(\ref{iik.eq3a})
 are reduced to Eqs. (\ref{Sph.eq1A}), (\ref{Sph.eq1Ba}),
 (\ref{Sph.eq1Bb}).

 Let us write Eq. (\ref{Sph.eq1A}) in the form
 \begin{equation}
 \label{Sph.eq1Aa}
 R =  2 \phi'.
 \end{equation}
 Substitution  of the expression for $R$  into  Eq. (\ref{Sph.eq1Ba})
 yields
 $V' = 4 \phi \phi',$ whence upon integration
 \begin{equation}
 \label{Sph.eq1Baa}
 V =  2 \phi^2 + A, \quad A = {\rm const.}
 \end{equation}
 Finally, substituting Eqs.  (\ref{Sph.eq1Aa}) and (\ref{Sph.eq1Baa})
 in Eq. (\ref{Sph.eq1Bb}) we obtain the following nonlinear second-order
 ordinary differential equation for $\phi(t):$
 \begin{equation}
 \label{Sph.eq5}
 \phi''+ 2\phi^3 + \left(A + \frac{f^2 + N^2}{2}\right)\phi = 0.
 \end{equation}

 Thus, we have arrived at the following result.
 \begin{thm}
 \label{spher:theor}
 The solutions of the system (\ref{iik.eq1a})-(\ref{iik.eq3a}) that are
 invariant with respect to the dilations and rotations with the
 generators $X_7$ and $X_9$ from (\ref{kov}) are given by
 \begin{align}
 & v = \frac{1}{f}\left[\left(2 \phi^2(t) + A + \frac{N^2 - f^2}{2}\right) x
 + 2 \phi'(t) z\right],\notag\\[1.5ex]
 & \rho =\frac{1}{g}\left[2 \phi' (t) x - \left(2 \phi^2(t) + A
 - \frac{N^2 - f^2}{2}\right) z\right], \label{Sph.eq4}\\[1.5ex]
 & \psi = (x^2 + z^2)\, \phi(t),
 \notag
 \end{align}
 where $\phi(t)$ is defined by the differential equation (\ref{Sph.eq5}) and $A$ is an arbitrary
constant.
 \end{thm}

 \subsection{Qualitative analysis of the invariant solution}

One can integrate Eq. (\ref{Sph.eq5}) once, e.g., upon multiplying
 by $2 \phi'$ and obtain
 \begin{equation}
 \label{Sph.eq5a}
 \phi'^2 + \phi^4 + \left(A + \frac{f^2 + N^2}{2}
 \right) \phi^2 = {\rm const.}
 \end{equation}
 We will analyze the behavior of the solutions to Eq.
 (\ref{Sph.eq5a}) under the assumption that
 the expression in the parentheses
 is a non-negative constant which we denote by
 $K:$
 \begin{equation}
 \label{Sph.eq5K}
 K = A + \frac{f^2 + N^2}{2}\,, \quad K \geq 0,
 \end{equation}
 and write Eq. (\ref{Sph.eq5a})  in the form
 \begin{equation}
 \label{Sph.eq6}
  \phi'^2 + \phi^4 + K \phi^2 = B^2, \quad B = {\rm const.,}
 \end{equation}
 or solving for $\phi':$
 \begin{equation}
 \label{Sph.eq7}
 \phi' = \pm \sqrt{B^2 - \phi^4 - K \phi^2}\,.
 \end{equation}

 Note that $\phi(t) = 0$ solves Eq. (\ref{Sph.eq5}). Let us turn
 to Eq. (\ref{Sph.eq7}). When $\phi$ is small, i.e. close to the trivial solution
 $\phi(t) = 0,$ then $$B^2 - \phi^4 - K \phi^2 \approx B^2$$ and hence $\phi'$
 is close to the constant value$$\phi' \approx \pm \,B.$$
 When $\phi(t)$ varies according to Eq. (\ref{Sph.eq5}), then
 $|\phi\,'|$ decreases since  $$B^2 - \phi^4 - K \phi^2 < B^2$$ when $\phi \not=
 0.$ We obtain $\phi' = 0$ when $\phi(t) = C_*,$ where
 \begin{equation}
 \label{Sph.eq8}
 C_*^2 = \frac{- K + \sqrt{B^2 + K^2}}{2}\,\cdot
 \end{equation}
 If $|\phi| > |C_*|,$ then $B^2 - \phi^4 - K \phi^2 < 0,$ and hence
 Eq. (\ref{Sph.eq7}) does not have a solution. We have arrived at the
 following significant results.
 \begin{thm}
 \label{behav:theor}
  Provided that the condition (\ref{Sph.eq5K}) holds,
  the solutions of Eq. (\ref{Sph.eq7})
 are bounded oscillating functions  $\phi(t)$ satisfying the
 condition
 \begin{equation}
 \label{Sph.eq9}
 - C_* \leq \phi(t) \leq C_*,
 \end{equation}
 where $C_*$ is the positive constant defined by Eq.
 (\ref{Sph.eq8}). In this notation, the invariant solution
 (\ref{Sph.eq4}) is written as follows:
  \begin{align}
 & v = \frac{1}{f}\left[\left(2 \phi^2(t) + K - f^2\right) x
 + 2 \phi'(t) z\right],\notag\\[1.5ex]
 & \rho =\frac{1}{g}\left[2 \phi' (t) x - \left(2 \phi^2(t) + K
 - N^2\right) z\right], \label{Sph.eq4K}\\[1.5ex]
 & \psi = (x^2 + z^2)\,\phi(t).
 \notag
 \end{align}

 \end{thm}
 %\subsection{Curvilinear wave beams}
 \begin{rem}
 The invariance of the solution  (\ref{Sph.eq4})
 with respect to rotations (rotational symmetry) means that
 it has the same values on any circle $$x^2 + z^2 = r^2$$ with a
 given radius $r.$ The invariance under dilations means that we can
 obtain the solution at any circle just by stretching the radius
 $r.$ According to Theorem \ref{behav:theor}, this solution is given by bounded
 oscillating functions.
 % and hence it is stable. We call them
 %\textit{spherical beams}.
 \end{rem}

 \section{Energy of the rotationally symmetric solution}
 \setcounter{equation}{0}

 The conservation of energy for Eqs. (\ref{iik.eq1a})-(\ref{iik.eq3a})
 has the form \cite{ibr-ibr09b}
 \begin{equation}
 \label{Sph:ener1}
 \frac{d}{d t} \int\!\int \left[v^2 + \frac{g^2}{N^2}\, \rho^2
 + |\nabla \psi|^2\right] dx d z  = 0.
 \end{equation}
 Hence, the \textit{energy density} is
 \begin{equation}
 \label{Sph:ener_dens}
 E = v^2 + \frac{g^2}{N^2}\,\rho^2 + |\nabla \psi|^2.
 \end{equation}

 For the rotationally invariant solution (\ref{Sph.eq4K}) we have
 \begin{equation}
 \label{Sph.eq4Kp}
 |\nabla \psi|^2 = 4 (x^2 + z^2)\,\phi^2(t).
 \end{equation}
 Substituting the expression (\ref{Sph.eq4Kp}) and the expressions (\ref{Sph.eq4K}) of $v$ and $\rho$
 in Eq. (\ref{Sph:ener_dens}) we obtain the following energy
 density for the invariant solution (\ref{Sph.eq4K}):
 \begin{equation}
 \label{Sph:ener_rot}
 \begin{split}
 E = & 4 \left(\frac{1}{f^2} - \frac{1}{N^2}\right) (x^2 - z^2)\big[\phi^2(t) +
 K\big] \phi^2(t)  + \left(f - \frac{K}{f}\right)^2 x^2\\[1ex]
 &
  + \left(N - \frac{K}{N}\right)^2 z^2
  +
 4 \left(\frac{1}{f^2} - \frac{1}{N^2}\right) xz\,\big[2 \phi^2(t) + K\big]
 \phi'(t).
 \end{split}
 \end{equation}\\[4ex]
 \null \hfill 22 April 2009

 \end{document}